\documentclass[12pt]{article}
\usepackage{psfig}
\title{Damage spreading and Lyapunov exponents in cellular
automata\footnote{This paper appeared (with modifications) 
in Phys. Lett. A {\bf 172} 34 (1992).}}
\author{F. Bagnoli$^{a,c}$,
R. Rechtman$^{b,}\thanks{On sabbatical leave from Departamento de F\'\i{}sica,
Facultad de Ciencias, Universidad Nacional Aut\'onoma de M\'exico,
Apdo. Postal 70-542, 04510 M\'exico D.F., Mexico}$ \hspace{.01in},
S. Ruffo$^{b,c}$\\
{}\\
${}^a\quad$~\small \it Dipartimento di Matematica Applicata,
Universit\`a di Firenze,\hspace*{\fill} \\
\small \hspace{.31in} \it  Via S. Marta 3, I-50139 Firenze,
Italy\hspace*{\fill}\\
${}^b\quad$~\small \it Dipartimento di Energetica, Universit\`a di
Firenze,\hspace*{\fill}\\
\small \hspace{.31in} \it Via S. Marta 3, I-50139 Firenze,
Italy \hspace*{\fill}\\
${}^c\quad$~\small \it Sezione I.N.F.N. and Unit\`a I.N.F.M. di Firenze,
Italy \hspace*{\fill}\\
}

\begin{document}
\maketitle
\newcommand{\D}{{\cal D}}
\newcommand{\M}{{\cal M}}
\newcommand{\B}{\hbox{\sf B}}

\begin{abstract}
Using the concept of the Boolean derivative we study damage spreading
for one dimensional elementary cellular automata and define their
maximal Lyapunov exponent. A random
matrix approximation describes quite well the behavior of ``chaotic''
rules and predicts a directed percolation-type phase transition.
After the introduction of a small noise
elementary cellular automata reveal the same type of transition.
\end{abstract}

\vspace*{\fill}

\section{Introduction}
The behavior of the distance between two configurations submitted to 
the same dynamics (damage spreading) is considered to be a good
tool to investigate the ergodic properties of the dynamics of 
discrete statistical models~\cite{Stauffer}. Although the relation
between these properties and ``chaotic'' behavior is still unclear,
there is an intuitive connection between ``chaos'' and damage
spreading on one side, and between a periodic attractor and damage
collapsing on the other. 
For continuous dynamical systems a positive
maximal Lyapunov exponent (MLE) implies chaotic motion. The MLE 
is roughly defined as the rate of the exponential
divergence of the distance between two initially close trajectories,
in the limit of long times and vanishing initial distances. In what
follows we show how Boolean derivatives may be used to define the MLE
of a cellular automaton.

A Boolean one-dimensional cellular automaton (CA) is a discrete dynamical
system defined on a lattice. The state of the system is represented by a
configuration  ${\bf x} = (x_1, \dots, x_i, \dots, x_L)$ of Boolean variables,
where $L$ is the size of the lattice. We always use
periodic boundary conditions ($x_{i+L}=x_i$). The time evolution of the
system 
is given by a deterministic Boolean function ${\bf F}$
\begin{equation}
\label{F}
	   {\bf x}(t+1) = {\bf F}\left({\bf x}(t)\right),
\end{equation}
which is in turn defined by a local, uniform rule $f$
\begin{equation}
\label{f}
	   x_i(t+1) =f(x_{i-r},\dots,x_i,\dots,x_{i+r});
\end{equation}
where $r$ is the range of the function $f$. 
There are $2^{2r+1}$ different CA of range $r$. In what follows we
restrict our study to elementary CA for which $r=1$, using Wolfram's
labeling convention~\cite{Wolf83}.

In the context of CA where time, space and dynamical variables are discrete,
we cannot extend directly the definition of Lyapunov
exponents
\cite{Wolf84,Wolf85,Packard,Jen,Shere,Wolf86}. Due to the finite
interaction range $r$ 
and to the finite number of states of the
variables $x-i$, 
the distance between two initially close configurations can
increase at most linearly for long times.

Instead of looking at the long time behavior of the  distance
between two configurations, we can use some hints from the theory of
continuous dynamical systems and study the local stability of a single
trajectory with respect to a small perturbation (a damage or defect in the
configuration). This defect can be readily recovered, or it can freeze
being replicated without change, or finally it can propagate increasing
the distance between the configurations.

In section~2, we show that the distance between two configurations
after the introduction of a defect is given by the Boolean Jacobian
matrix of the evolution function ${\bf F}$.
While in the actual evolution of the automaton the defects can interact
and annihilate themselves, we are interested in the stability of a single
trajectory, and we restrict to the case of non-interacting defects
which is equivalent to consider a product of Boolean Jacobian
matrices on a trajectory. For the elementary CA it is a Jacobi matrix
with elements equal to zero or one on the three main diagonals.  This
in turn suggests a relation with the product of random matrices of
the same type.
The MLE of the product of these random matrices shows a transition related
to that of directed percolation~\cite{Hammer,Kinzel}.

As reported in section 3, the results of simulations of ``chaotic'' CA
(whose space-time patterns, starting from a random configuration are
disordered and aperiodic)  agree quite well with the predictions of
the random matrix approximation. Adding a noise on the evolution of the 
automaton, our approach reveals the existence of a
transition in the space of CA rules from a ``frozen phase'' where damage does
not spread, to a phase where damage spreads locally with a positive MLE close to
the one given by the product of random matrices.

\section{Boolean derivatives, defects and random matrices}

We are interested in the local stability with respect to a small
perturbation,
of the time evolution (\ref{F}) of the configuration $\bf x$.
Let us denote a defect at site $i$ the configuration ${\bf z}^{(i)}$
with elements $z_j=\delta_{i,j}$, \mbox{$j=1,\dots,L$} and $\delta_{i,j}$
the usual Kronecker symbol. The configuration
${\bf y}(t) = {\bf x}(t) \oplus {\bf z}^{(i)}$ differs from ${\bf x}(t)$
only at site $i$, where the XOR operation ($\oplus$) is performed site by
site.
Depending on $\bf F$ and on the configuration {\bf x}, the defect
${\bf z}^{(i)}$
can originate in one time step up to three defects in sites $i-1$, $i$ and
$i+1$. We write
$$
{\bf x}(t+1) \oplus {\bf y}(t+1) =
        F^{\prime}_{i-1,i} \wedge {\bf z}^{(i-1)}
 \oplus F^{\prime}_{i,i} \wedge {\bf z}^{(i)}
 \oplus F^{\prime}_{i+1,i} \wedge {\bf z}^{(i+1)};
$$
where $F^{\prime}_{i,j} = 0,1$ and 
the symbol $\wedge$ represents the AND operation, to be performed between 
the number $F'_{i,j}$ and each element of the configuration ${\bf z}$.
The quantities 
$$
F^{\prime}_{i,j} = \frac{\partial x_i(t+1)}{\partial x_{j}(t)}.
$$
are the elements of the Boolean Jacobian matrix
${\bf F}^{\prime}$ of ${\bf F}$; they are expressed by the Boolean derivatives
of the local evolution rule $f$ of eq. (\ref{f}) \cite{Vichniac,Boolean}.
For instance
$$
\frac{\partial x_i(t+1)}{\partial x_{i+1}} = 
f(x_{i-1},x_i,  x_{i+1}) \oplus f(x_{i-1},x_i, ,x_{i+1}\oplus 1).
$$
As  $f$ has range 1, $\partial x_i(t+1)/\partial x_{j}$ vanishes if 
$|i-j| > 1$, and $\bf F^{\prime}$ is a Jacobi matrix whose elements are
zero or one.
If the local evolution rule is expressed in terms of AND and XOR
operations (ring sum expansion), the Boolean derivatives extract the linear
part of $f$.

We are interested in the limit of a small initial perturbation
to a given trajectory. This limit corresponds in discrete dynamics to
the presence of only one point defect. If, during the evolution,
$m$ defects appear, we consider $m$ of replicas of the
system and assign one of the defects to each one. We
indicate with $N_i(t)$ the number of replicas carrying the defect
${\bf z}^{(i)}$ at time $t$. If for instance we start at time zero
with only one defect at some site $i$ ($N_i(0)=1$),
and the rule allows the spreading of the defects to the sites of the
neighborhood at each time step, at $t=1$ \mbox{$N_{i-1}=N_i=N_{i+1}=1$},
at $t=2$ $N_{i-2}=N_{i+2}=1$, $N_{i-1}=N_{i+1}=2$, $N_i=3$, etc.

The time evolution of the number of defects at site $i$ is given by
$$
    N_i(t+1) = \sum_j  F^{\prime}_{i,j}(t) N_j (t),
$$
or, in matrix form,
\begin{equation}
\label{defects}
    {\bf N}(t+1) = {\bf F}^{\prime}{\bf N}(t), 
\end{equation}
where the elements of ${\bf F^{\prime}}$ are now interpreted as integer
numbers. It is worth noting
that $N_i(t)$ is also the number of paths in defect space that reach the
site $i$ at time $t$ starting from any defect at time $t=0$.

We define the finite-time MLE $\lambda(T)$ of the
mapping (\ref{defects}) as
\begin{equation}
        \label{lyap}
        \lambda(T) = \frac{1}{T} \sum_t \log \eta(t).
\end{equation}
where the local expansion rate of defects $\eta$ is defined as
\begin{equation}
\label{eta}
\eta(t) = |{\bf N}(t+1)| / |{\bf N}(t)|,
\end{equation}
and $|{\bf N}| = \sum_i N_i$.  
In the following
$\lambda(\infty)$ will be denoted simply by $\lambda$.
This definition is meaningful because the number $|{\bf N}(t)|$ can diverge
exponentially.

If $\lambda < 0$ the number of defects (the damage) decreases exponentially
to zero, while if $\lambda > 0$ the damage spreads. Let us give some simple
examples. Rule 0 that maps all the configurations to the configuration
$\{0\}^L$, has $\lambda = -\infty$ because the Jacobian is zero.
The ``chaotic'' rule 150
has $\lambda= \log 3$, because all its
Boolean derivatives are equal to one. A marginal case is rule 204, for which
${\bf F}^{\prime}$ is the identity matrix and $\lambda = 0$. The derivatives
of the 88 ``minimal'' elementary CA may be found in ref.~\cite{Vichniac}.

In the spreading case, a reasonable approximation
to the dynamics of defects
(\ref{defects}) consists in substituting
the deterministic matrix ${\bf F}^{\prime}$ with a random matrix of the
same form. We therefore consider the product of
random tridiagonal matrices ${\bf M}(p)$ having a fraction $p$ of
elements on the three principal diagonals equal to one.
The quantity $p$ is interpreted as the geometric mean $\mu$ of
the derivative on the CA configuration for large $T$, i.e.,
$$
\mu(T) = \left(\prod_{t=1}^T \mu(t)\right)^{1/T}
$$ 
and
$$
    \mu(t) = \frac{1}{3L} \sum_{i=1}^L \sum_{k=-1}^1 F^{\prime}_{i,i+k}.
$$

The evolution of the number of defects in the random matrix
approximation defines a directed bond percolation problem with control
parameter $p$, assuming that a site at location $i$ at time $t$ is
``wet'' if $N_i(t) > 0$. Observe that $N_i(t)$ gives
the number of directed paths that reach site $i$ at time $t$
inside the  percolating cluster. Therefore we expect a second--order
phase transition at $p=p_c$ with order parameter the density of wet
sites $\rho(t)$.

We have first localized the percolation threshold at $p_c=0.441(1)$
(where the number in parenthesis is the error on the last significant
digit) by looking at the asymptotic behavior of the order parameter. Then,
starting with an initial condition where all the sites are wet,
we have verified that $\rho(t) \sim t^{-\beta/\nu_\parallel}$ at $p_c$
with $\beta/\nu_\parallel = 0.155(3)$ the usual exponents of directed
percolation. 

The results of the random matrix approximation are reported in 
Fig. 1, where the curve shows the dependence of the MLE $\lambda$
for the product of random matrices ${\bf M}(p)$ as a function of $p$. 
For $p<p_c$ 
$\lambda=-\infty$. At $p_c$, for sufficiently large $T$,
$$
\lambda (T) = \lambda + a T^{-\chi}
$$
with $\lambda=0.237(2)$, $\chi = 0.68(4)$ and $a > 0$. This shows that
at the critical point the number of walks on the percolation cluster
grows exponentially with time, with an effective coordination
$\exp(\lambda)$. The exponent $\chi$, which is not usually defined
in percolation,
might be related to the critical exponents for directed walks~\cite{Kardar}.
The data at the percolation threshold were obtained by
letting a $10^4 \times 10^4$ random tridiagonal matrix evolve during
$4 000$ time steps for $30$ realizations.

\begin{figure}
\centerline{\psfig{figure=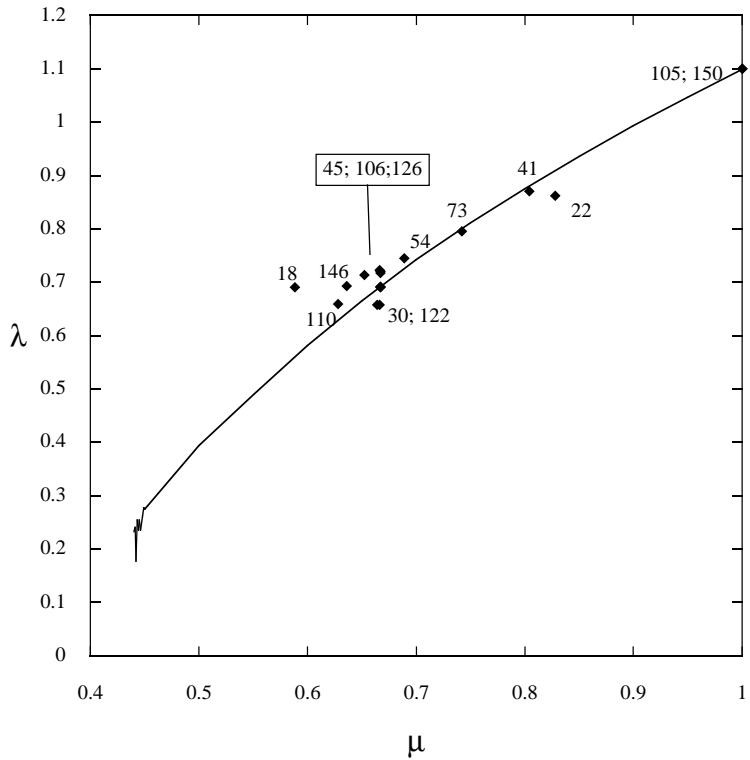,width=10cm}}
\caption{
The curve shows the MLE $\lambda$ of a
random tridiagonal matrix as a function of $p$.  The diamonds show the
asymptotic value of $\lambda$ for ``chaotic'' CA.
Results for CA were obtained by letting each automaton evolve
during $5 000$ time steps on a $512$ sites lattice and
$\alpha_0 =0.5$.
}
\end{figure}

We obtain a mean field approximation replacing $\bf M$ with a constant
tridiagonal matrix with elements equal to $p$. The corresponding MLE 
is $\lambda=\log 3p$ which is positive for $p \ge 1/3$.
This approximation agrees well with numerical
simulations for $p\ge p_c$, with a maximum deviation of $18\%$ at $p=p_c$.  

In the numerical calculation of the Lyapunov exponent $\lambda$ one needs
to renormalize ${\bf N}(t)$~\cite{Benettin}. This is impossible if
${\bf N}$ is defined over integers. However, since the Lyapunov exponents
are independent of the choice of the norm~\cite{Oseledec},
we let ${\bf F}^{\prime}$ (or ${\bf M}$) act on some abstract ``tangent''
space in {\sf R}$^L$, using the usual Euclidean norm. Applying standard
methods one obtains the Lyapunov exponents related to the exponential
divergence of the norm of the product of
$\sqrt{{\bf F}^{\prime^{\dagger}} {\bf F}}$ where $\bf F^{\prime^{\dagger}}$
is the Hermitian conjugate of ${\bf F}^{\prime}$. The standard definition of
the MLE coincides with the one of eq.(\ref{lyap}).

\section{Elementary cellular automata}

We computed the mean number of ones $\mu(T)$ in the Jacobian matrix and the
finite-time MLE $\lambda(T)$ for all the 88 ``minimal'' elementary CA
for $L=256$ and $L=512$ and $5 000 \leq T \leq 15 000$
starting from random initial configurations with a fixed fraction
$\alpha_0$ of live sites, $\alpha_0 =L^{-1} \sum x_i(0)$.
The quantities $\mu(T)$ and $\lambda(T)$ are generally already asymptotic for
$T \sim 5 000$; moreover they show a very
weak dependence on $\alpha_0$ for $0.2 \leq \alpha_0 \leq 0.8$ (only
rules 6, 25, 38, 73, 134 and 154 vary more than $10\%$ but less than $20\%$).

\vspace{.5cm}
We note that
\begin{itemize}
\item{i)} CA with constant ${\bf F}^{\prime}$ independent of the
configuration (e.g. rules 0, 15, 60, 90, 150, 204) have
$\lambda = \log 3 \mu$ with $\mu= 0$, $1/3$, $2/3$ or $1$.
\item{ii)} CA for which all configurations are mapped to a homogeneous state
(e.g. rules 8, 32, 40, 136) have $\lambda=-\infty$. The control parameter
$\mu$ is zero. These are {\em class 1} CA in Wolfram's classification
\cite{Wolf84}.
\item{iii)} ``Chaotic'' {\em class 3} CA with nonconstant
${\bf F}^{\prime}$ (e.g. rules 18, 22, 30, 41, 45, 106, 110, 122, 126,
146) have $\mu > p_c$, $\lambda > 0$ and the damage spreads.
\end{itemize}

The values of MLE for the ``chaotic'' CA 
of cases {\em i)} and {\em iii)} the value of
$\lambda$ agrees well with the random matrix approximation, as shown
in Fig.~1. This is also trivially true for rules of case {\em ii}.

\begin{figure}
\centerline{\psfig{figure=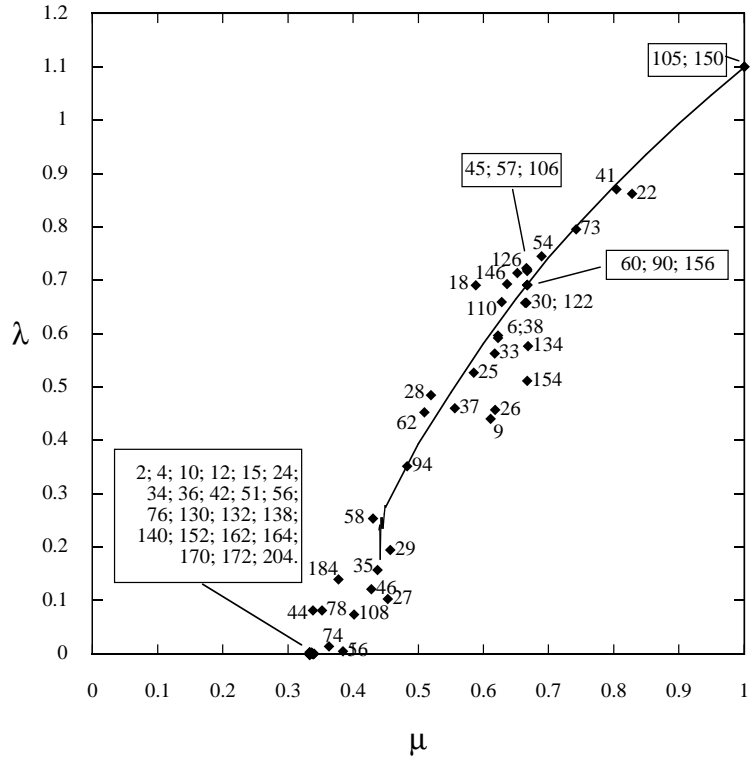,width=10cm}}
\caption{
The curve is the same as the one shown in Fig. 1.
The diamonds show the values of $\mu$ and $\lambda$ for all the minimal
CA with $\lambda \geq 0$.
}
\end{figure}

For the automata whose evolution leads to a nonhomogeneous periodic
space pattern (e.g. {\em class 2} CA), $\lambda$ is the logarithm of
the largest eigenvalue of the product of the Jacobian matrices over the
periodic state. The measured value of $\lambda$ is always non negative. This
suggests that the asymptotic state is unstable ($\lambda > 0$) or
marginally stable ($\lambda = 0$). One can think that  
the ``freezing'' of the evolution 
occurs because
there are no ``close'' configurations which can be used as an intermediate
state towards a more stable state. Therefore, we 
``heated'' the evolution
by exchanging the states of a small number $s$ of pairs of randomly chosen
sites at each time step. We observe that all the rules which had $\mu <
1/3$ 
(the critical point in the mean field approximation)
relax to $\mu=0$ and
$\lambda = -\infty$, those with $1/3 < \mu < p_c$ go to $\lambda = 0$ and
the rules with $\mu > p_c$ approach the random matrix result. 
In Fig. 2, we show the location of all the 88 minimal elementary rules
with noise in the $(\mu,\lambda)$ diagram, starting with $\alpha_0 = 0.5$.

After the introduction of noise, the CA rules can be divided
roughly in three groups, according to the value of their MLE.
In the first group, with $\lambda=-\infty$ and $\mu=0$ we find  all
class 1 CA (rules  0, 8, 32, 40, 128, 136, 160 and 168) and some class
2 CA (rules 1, 3, 5, 7,
11, 13, 14, 19, 23, 43, 50, 72, 77, 104, 142, 178, 200 and 232). 
Rules 50, 77 and 178 show very long transients of the order of $15,000$
time steps. Rule 232, a majority rule, illustrates well a typical
behavior. Configurations $\{0\}^L$ and  $\{1\}^L$
are fixed points for this rule. A single defect in these
configurations is recovered in one time step. On the other hand, an
arbitraty initial configuration will give in a few time steps a pattern of
strips. By adding a noise as described above, the borders of the
strips perform a sort of random motion, thus allowing their merging.
Finally, one of the two fixed points is reached, according to
the initial density of the configuration. 

In the second group we find the ``chaotic'' {\em class 3} rules.
The values of
$\mu$ and $\lambda$ are slightly affected by the noise. They have
$\mu > p_c$ and are close to the curve of
the  random matrix approximation for $\lambda$. We also find CA which are
not {\em class 3} (rules 6, 9, 25, 26, 27, 28, 29, 33, 37, 38, 54, 57,
62, 73, 94, 134, 154, and 156) but show local damage spreading.

The third group contains
exceptions to the random matrix approximation. This occurs
for rules with a value of $\mu$ without noise close to $1/3$
(CA 15, 51, ...204)
or $p_c$ (CA 1,3,5,11,14,43,142,24,44,46,56,74,108 and 152)
Contrary to the prediction of the random matrix
approximation ${\bf N}$ does not vanish.
Moreover, it should
be remarked that
CA with a large number of conserved additive
quantities (rules 3, 4, 5, 10, 12, 15, 34, 42, 51, 76, 138, 140, 170, 200
and 204)~\cite{Takesue} have indeed $\mu=1/3$ and $\lambda=0$.

In this letter we have shown how the maxumum Lyapunov exponent
can be defined for CA using the Boolean derivative.
A positive Lyapunov exponent is associated to local damage spreading and on
the other hand reflects the exponential growth of paths on the
percolation clusters. For CA with $0<\mu<p_c$ and a positive Lyapunov exponent
the introduction of a small noise produces the collapse to
$\lambda=0$ or $\lambda=-\infty$. A random matrix model is
directly suggested by the CA dynamics and displays a directed percolation
phase transition. A somewhat similar phase transition is observed in
CA rule space in the presence of a small amount of noise.
The extension of our definition of Lyapunov exponent to other discrete
systems, and possibly to probabilistic dynamics will be the
subject of future investigations.

\section*{Acknowledgments}
This work sprang from a discussion of one of the authors (S.R.) with
G. Vichniac. We are grateful to R. Bulajich, R. Livi and A. Maritan
for fruitful discussions and suggestions. This work was partially
supported by CNR of Italy, CONACYT and DGAPA-UNAM of Mexico.

\end{document}